\documentclass[12pt]{tcibook}
\usepackage{fancyhea}
\usepackage{work}
\usepackage{bm}       
\usepackage{graphicx}
\usepackage{multirow}
\usepackage{lineno}
\usepackage{threeparttable}
\usepackage[normalem]{ulem}
\usepackage{color}
\usepackage[colorlinks = true,
            linkcolor = blue,
            urlcolor  = blue,
            citecolor = blue,
            anchorcolor = blue]{hyperref}

\def\beq{\begin{equation}}
\def\eeq#1{\label{#1}\end{equation}}
\def\eeqn{\end{equation}}

\newenvironment{Eqnarray}%
   {\arraycolsep 0.14em\begin{eqnarray}}{\end{eqnarray}}
\def\beqa{\begin{Eqnarray}}
\def\eeqa#1{\label{#1}\end{Eqnarray}}
\def\eeqan{\end{Eqnarray}}

\let\bar=\overbar

\def\lsim{\mathrel{\raise.3ex\hbox{$<$\kern-.75em\lower1ex\hbox{$\sim$}}}}
\def\gsim{\mathrel{\raise.3ex\hbox{$>$\kern-.75em\lower1ex\hbox{$\sim$}}}}

\def\del{\partial}
\def\Dslash{\not{\hbox{\kern-4pt $D$}}}
\def\dslash{\not{\hbox{\kern-2pt $\del$}}}
\def\pslash{\not{\hbox{\kern-2pt $p$}}}
\def\ETmiss{\not{\hbox{\kern-4pt $E$}}_T}

\def\Dlr{\mathrel{\raise1.5ex\hbox{$\leftrightarrow$\kern-1em\lower1.5ex\hbox{$D$}}}}

\def\MSB{{\bar{M \kern -2pt S}}}
\def\msb{{\bar{\scriptsize M \kern -1pt S}}}

\def\drb{{\bar{\scriptsize D \kern -1pt R}}}

\def\authorlist#1#2{
    \vskip 0.4in
\begin{center}\begin{large} {\bf  #1 } \end{large}
    \vskip 0.2in
              #2
     \vskip 0.2in
   \end{center}
}

\begin{document}


\pagenumbering{roman}

\parindent=0pt
\parskip=8pt
\setlength{\evensidemargin}{0pt}
\setlength{\oddsidemargin}{0pt}
\setlength{\marginparsep}{0.0in}
\setlength{\marginparwidth}{0.0in}
\marginparpush=0pt


\pagenumbering{arabic}

\renewcommand{\chapname}{chap:intro_}
\renewcommand{\chapterdir}{.}
\renewcommand{\arraystretch}{1.25}
\addtolength{\arraycolsep}{-3pt}

\setcounter{chapter}{0} 


\chapter{Experimental Algorithm Parallelization}

\authorlist{Conveners: G. Cerati, K. Heitmann, W. Hopkins}
   {Contributing Authors: J.~Bennett, T.Y.~Chen, V.V.~Gligorov, O.~Gutsche, S.~Habib, M.~Kortelainen, C.~Leggett, R.~Mandelbaum, N.~Whitehorn, M.~Williams}

\begin{center}
\today
\end{center}

\section{Executive Summary}

The substantial increase in data volume and complexity expected from future experiments will require significant investment to prepare experimental algorithms. These algorithms include physics object reconstruction, calibrations, and processing of observational data. In addition, the changing computing architecture landscape, which will be primarily composed of heterogeneous resources, will continue to pose major challenges with regard to algorithmic migration. Portable tools need to be developed that can be shared among the frontiers (e.g., for code execution on different platforms) and opportunities, such as forums or cross-experimental working groups, need to be provided where experiences and lessons learned can be shared between experiments and frontiers. At the same time, individual experiments also need to invest considerable resources to develop algorithms unique to their needs (e.g., for facilities dedicated to the experiment, such as online farms), and ensure that their specific algorithms will be able to efficiently exploit external heterogeneous computing facilities. Common software tools represent a cost-effective solution, providing ready-to-use software solutions as well as a platform for R\&D work. These are particularly important for small experiments which typically do not have dedicated resources needed to face the challenges imposed by the evolving computing technologies. Workforce development is a key concern across frontiers and experiments, and additional support is needed to provide career opportunities for researchers working in the field of experimental algorithm development. Finally, cross-discipline collaborations going beyond high-energy physics are a key ingredient to address the challenges ahead and more support for such collaborations needs to be created. This report targets future experiments, observations and experimental algorithm development for the next 10-15 years. In summary, experimental algorithms will be faced with a range of challenges and the need to invest in a range of possible solutions:
\begin{itemize}
    \item {\bf Challenges}
    \begin{itemize}
        \item Substantial increase in data volume and complexity from future experiments
        \item Finite amount of compute resources
        \item Changing computing architecture landscape and algorithm migration
        \item Support for small-scale experiments 
    \end{itemize}
    \item {\bf Recommended Investments}
    \begin{itemize}
        \item Development of portable solutions and common software tools
        \item Work force development 
        \item Career opportunities in the area of experimental algorithm developments
        \item Support of cross-discipline collaborations going beyond high-energy physics 
    \end{itemize}
\end{itemize}

\section{Introduction}

The computational landscape has changed dramatically over the last 15 years. In 2008, the petascale barrier was broken for the first time by the Roadrunner supercomputer at Los Alamos, a hybrid machine with CPUs and local computational accelerators. Since then, hardware-accelerated designs have dominated High Performance Computing (HPC) architectures. At the same time, the demand from the high energy physics community for computing resources for experiments continued to steadily increase, but with a focus on high-throughput computing (HTC), in many cases provided by grid and local computing clusters. As the hardware evolution continues, there is a question of how the HEP community should address possible opportunities in both of these computational paths. Given the importance of the questions, reports have been written by different groups, e.g. Ref.~\cite{HEPSoftwareFoundation:2020daq} discusses tools and software relevant for HL-LHC, Ref.~\cite{HEPcompWP} includes also discussions about DUNE and Belle-II, and Ref.~\cite{2019arXiv190505116B} provides comprehensive information for the field of astronomy.

Experimental algorithms include algorithms that calibrate measurements, process observational data to reconstruct physics objects (e.g., electrons, photons, muons, etc). These algorithms are developed by large collaborations over long periods of time and are often embedded in complex software ecosystems, and then further developed and maintained by the collaborations. Due to this complexity, the evolution is usually rather slow and often cannot keep pace with developments in the computing hardware landscape and associated programming models. Also, experiments typically have access to a broad range of hardware in operation at any given time, so they can't optimize their software just for the latest hardware. These three challenges, rapidly evolving hardware and programming models, ever increasing demands on computing resources, and relatively slow development cycles of experimental software by large teams, has led to several desiderata:
\begin{itemize}
    \item Performance portability libraries to support multiple computing architectures without significantly impeding performance and minimizing platform-specific code;
    \item Enabling the use of HPC resources in addition to the traditional HTC approach to increase the pool of available computational resources;
    \item Construction of computational frameworks that allow for collaborative software development and code reuse;
    \item Sophisticated new algorithmic approaches and integration of AI/ML methods to tackle the analysis of complex data sets and to reduce computational costs.
\end{itemize}
In addition to the challenges that affect all HEP programs, experimental and observational sciences exhibit a major difference in that energy and intensity frontier experiments carry out controlled experiments while in the cosmic frontier the experiments are observations of the Universe, whose modeling is still affected by large uncertainties. Therefore, experimental algorithms for the Energy, Neutrino Physics and Rare Processes and Precision Frontiers, while still complex, present events and signals that are better defined than in the Cosmic Frontier, where signals are often more subtle and difficult to protect from systematics. 

The typical model for experimental parallelization work relies on funding for software and computing R\&D projects, targeting investigation of new technologies for HEP. R\&D projects however often result in demonstrations of applicability, thus not guaranteeing support for additional steps of vital importance such as integration in the experiments' software and maintenance over the experiment lifetime.

A common theme across all frontiers is the design of experimental algorithms and efforts to make them run efficiently on modern parallel architectures, and is typically separate from the design process of the detectors. As a result, choices in terms of detector technology or configuration may be suboptimal from the point of view of the computational efficiency of algorithms used to process the data they collect. A unified design of the detector and the associated software may lead to a cost optimization when both the hardware and computing costs are taken into account~\cite{Krutelyov:2016wxl}.

In this report, we will first discuss challenges and opportunities across the physics frontiers that arise from the new HPC architectures, including the role of software frameworks. Next we will discuss experimental algorithms in the context of AI/ML methods. We will then highlight challenges and opportunities unique to each of the physics frontiers. Finally, we provide a set of recommendations.

\section{Experimental Algorithm Developments on HPC Architectures: Challenges and Opportunities}


The computational resource requirements have grown rapidly in recent years and will continue to grow in the future. Experiments are expected to generate significantly more data and new experiments across the frontiers are beginning to demand large computing allocations to enable data processing and analysis. This increase in demand poses challenges to providing sufficient resources across the frontiers. In the last few years, HPC resources have been suggested to aid in satisfying the needs of the community. They provide an outstanding opportunity to tap into new compute resources and provide relief for the experiments. However, in particular for experimental algorithm development, they pose key technological challenges that need to be overcome to enable the usage of these more powerful and more diverse computing resources and architectures. There are examples in the literature that show successful use of HPC resources, focused on specific tasks, e.g. optimal parallel execution of algorithms~\cite{Childers:2015tyv}, but in general, these implementations require a significant commitment from developers. Challenges in preparing experimental algorithms for HPC resources include the effort and expertise required to rewrite complex experimental algorithms to make use of new types of architectures (e.g., GPUs) and preparing these algorithms to run on multiple types of processing units (e.g., CPUs, FPGAs, GPUs). It is very likely that such focused efforts will continue in the future but will be restricted to a small number of algorithms that can be easily adapted to HPC resources.
To address these challenges, investment in workforce development is crucial. Members of the community need to be trained in how to optimally use and program these resources. It will also require training in algorithm design and implementation to take advantage of new hardware solutions.

A more practical way to enable the use of HPC resources more broadly, is via the implementation of experimental algorithms using portability tools. This often means sacrificing optimal performance but reduces the effort needed to optimize algorithms for each architecture separately. There are still many questions that need to be answered, e.g. which tool to use and how much performance is lost, but steady progress is being made in this direction.

Finally, software frameworks play an important role with regard to experimental algorithm development as well. Usually, the algorithms are deeply embedded in such frameworks and they need to support different platforms.

It's important to note that, while experimental algorithms are being optimized for modern architectures, experiments should develop a roadmap to advance their computing needs, techniques, and facilities to the evolving computing environment. In this respect, a notable example is the trend to bring the CPU compute pattern closer to the one of GPUs, for instance with respect to memory access, or even to physically share the same socket. Such developments may prove beneficial for experimental workflows with low compute intensity that will thus reduce overheads from memory access.

In the following, we discuss first the parallelization efforts for specific algorithms, then portability solutions and finally the role of software frameworks in more detail. All of these approaches aim in different ways to enable the use of new architectures and have different advantages and disadvantages.

\subsection{Parallelization of Specific Algorithms}

In recent years many R\&D projects or efforts within experiments have focused on the re-design of existing algorithms, or the development of new algorithms, that specifically target a given architecture or set of architectures. These include both algorithms that are executed in the offline reconstruction workflows, as well as algorithms that are part of software triggers. Some examples, divided by frontier, are listed in Sec.~\ref{sec:algsfronts}.

This approach presents clear advantages as well as several disadvantages. The main advantage is that it allows the experiment to focus on and speed up specific algorithms which constitute the main consumers of the total computing time. On the other hand, these efforts require significant work by a team of physics and software experts, which cannot necessarily be directly applicable to other algorithms. Another advantage is that, if the algorithms are optimized for the architecture on which most of the experiment's processing takes place (e.g. an online farm for triggering, or a T0/T1 site for offline, or a specific HPC center), the benefits in terms of throughput are maximized. However, these speedups are not necessarily portable to other computing sites, which may feature different architectures.

In summary, this is likely the optimal option for specific time consuming algorithms, but it is not a viable way for parallelization of full workflows that need to run on heterogeneous computing centers since these workflows tend to be comprised of hundreds of algorithms.

\subsection{Portability Solutions}
\label{sec:portability}
Portable software solutions that enable the execution of experimental algorithms on a range of different architectures without or with very minimal changes to the code infrastructure are essential for the continuing scientific success of our experiments. Additionally, portability solutions or standardized parallelization schemes (e.g., including parallelization in the C++ standard) will facilitate the training of the HEP workforce to make use of future heterogeneous computing resources. In this section we summarize the findings of the Snowmass White Paper "Portability: A Necessary Approach for Future Scientific Software"~\cite{Bhattacharya:2022qgj}. 

In the past and mostly still today, experimental algorithms for HEP have been designed to efficiently run on x86 platforms. The codes literally ran (and still run) everywhere, from computing centers to our own laptops without requiring any tweaks to the codes themselves. With the advent of new hardware solutions to provide increased performance this scenario has changed. The heterogeneity of the available computing hardware requires tailored solutions to achieve high performance. Given the large code and user bases for HEP experimental algorithms, implementing such tailored solutions is often very difficult and sometimes impossible given the effort required.

As mentioned before, investing effort on platform-specific parallelization of targeted algorithms is often not cost-effective. A better strategy, championed by the High Energy Physics Center for Computational Excellence (HEP-CCE)~\cite{Bhattacharya:2022qgj}, involves performance studies of candidate portability tools (Kokkos, SYCL, Alpaka and others) on HEP-relevant test beds.

The HEP-CCE focuses on the evaluation of portable solutions and aims to provide guidance to the broader community to tackle this challenge. As part of the HEP-CCE, the Portable Parallelization Strategies project has been initiated which consists of the following components:
\begin{itemize}
    \item Identification of plausible portability solutions, including Kokkos, SYCL, Alpaka, \\OpenMP/OpenACC and $\mathtt{ std::execution::parallel}$
    \item Identification of a small number of HEP relevant test beds
    \item Definition of a set of metrics to evaluate the different solutions
\end{itemize}
This set-up is extremely useful for evaluating different approaches for real-life applications. The approach not only provides the community with guidance about the pros and cons for the different solutions but also derives a set of metrics used to make a decision on what tool to adopt. The paper provides two major take-away messages: (i) workable portability solutions are extremely important to ensure the continued scientific success of our experiments by enabling access to a wide range of computing resources; (ii) approaching this problem as a community rather than separately for each experiment is the best strategy to deliver optimal technical solutions without duplication of effort.

The report~\cite{Bhattacharya:2022qgj} lists fourteen metrics to evaluate different solutions, documented on GitHub\footnote{https://hep-cce.github.io/Metric.html}. They include, e.g., ease of learning, feature availability and long term sustainability and code stability. While all of these metrics are very relevant, the overriding consideration is the overall cost that would arise, including hardware and electricity, if we are not able to exploit the new architectures available to the community. Balancing all these considerations and developing a roadmap that describes suitable and affordable approaches going forward to ensure that computational resources are available and accessible to process and analyse the very large data sets that will be collected over the next decades is of utmost importance. 

\subsection{Common Tools}
\label{sec:comtools}

Several examples exist of tools that provide functionalities for more than a specific experiment. These common tools include software frameworks, such as {\it art}~\cite{art} or Gaudi~\cite{gaudi}, as well as shared repositories for algorithms, such as LArSoft~\cite{Snider:2017wjd} and ACTS~\cite{Ai:2019kze,Gumpert_2017}. By collecting experiment-independent implementations of algorithms through generic interfaces, these tools provide an infrastructure that gives access to knowledge and functionalities that is available to a broad set of experiments. For instance, LArSoft develops and supports a shared base of simulation and reconstruction software across liquid argon time projection chamber (LArTPC) neutrino experiments, while ACTS is a toolkit for charged particle track reconstruction in high energy physics experiments. 

In order to make use of these tools, experiments typically need to implement detector-specific descriptions (such as geometry, material, magnetic field) and couple them with the generic interface provided by the tool. This can significantly decrease the cost and time to results for new efforts, especially for small experiments~\cite{FASER:2022yqp} that may not have the resources to build from ground-up their entire software stack in the ever-evolving current computing landscape. In addition to providing access to ready-to-use software solutions, shared tools also serve as platforms for R\&D developments, thus providing common benchmarks and data sets for evaluation of different approaches~\cite{TrackML}, and give early access to innovative solutions to a large spectrum of experiments. R\&D work accessed through these tools may span the usage of AI methods, algorithm parallelization, and deployment on HPC resources, which can significantly reduce the required workload for smaller experiments. The main drawbacks are that experiment-specific tuning of algorithms will still be needed to be performed by each experiment, and specific handles available to a given experiment (e.g. due to innovative detector features) may not be immediately supported by the common tool. In addition, usage of common tools requires a buy-in to the choices of software architecture and conventions of the tool, which need to be carefully evaluated by each experiment and use-case.

Despite these drawbacks, common tools can represent the most cost-effective solution for the needs of many experiments. Their development thus needs to be promoted, and support for them needs to be provided in the long term to guarantee that efficient usage of the tools can be made by many experiments over their full life cycle.

\subsection{Role of Software Frameworks} 

Experimental algorithms are typically run as part of data processing workflows executed within the software framework of the experiment. It is therefore required that the desired level of parallelism, and its specific implementation within the algorithms, are supported at the framework level. In addition to parallelism on multi- or many-core CPUs, frameworks also need to support execution of algorithms on heterogeneous platforms, including for instance CPUs, GPUs, and FPGAs. The evolution of software frameworks needed to face these challenges is described in Ref.~\cite{Jones:2022ycw}. 

Traditionally, collider based HEP software frameworks partition the data in independent units (events) corresponding to a single detector readout. This requirement needs to be revisited for several reasons. On one side, for collider experiments the data from a single event is not enough to keep computational accelerators fully occupied, so concurrent processing of multiple events is needed. On the other side, large homogeneous detectors such as DUNE produce very large size raw event data, such that the memory needed to process one event exceeds the nominal per-core value, so dividing the event data in sub-units may allow for better optimizations.

In addition, modern frameworks typically organize data as arrays of structures (AoS) where structures (or classes) make extensive use of object-oriented features such as multiple levels of polymorphic inheritance in their event data models (EDMs). This ensured a consistent usage of conceptually similar data objects (e.g. pixel and silicon strip hits) through unique interfaces in C++/CPU programs, while at the same time preserving the specific features of each object type.
However, computational accelerators prefer flat structures (e.g. structures of arrays, or SoA) where threads access sequential or regularly stepped memory locations and cannot use objects of the type mentioned above efficiently. 
Therefore data objects need to be redesigned for performant access on accelerators, either by modifying the EDM, thus requiring propagating changes to a large amount code, or by converting them when they are transferred at run-time to and from the accelerator, thus potentially becoming a limiting factor in terms of performance. Further development is needed in this area to find the best trade-offs between performance and usability, requiring investigations of alternative EDMs, intermediate event data storage layers for efficient processing at HPC~\cite{Bashyal:2022dul}, or innovative methods for faster data conversions.

Optimal execution of specific algorithms or at specific computing centers may require compiling the code with specific compiler versions or directives. This approach runs the risk of leading to many variants of an experiment’s code base which will have to be managed and distributed, which is impractical given the O(100) GB size of the full software stack builds. Mitigation strategies need to be explored and implemented, and may include fat binaries, selective optimizations, or on-site builds. Additionally, physics validations will need to be carried out for all configurations, which is a non-trivial task. 

As an example for the degree of parallelism available in software frameworks, we consider here the CMS experiment's framework (CMSSW)~\cite{jones:2006}, which has been on the front line in terms of enabling parallelism within its code. Similar features have been implemented or are being considered by other software frameworks, including Gaudi and {\it art}. As of 2022, CMSSW supports the following levels of concurrency:
\begin{itemize}
\item Independent framework modules (algorithms) that process independent subsets of the data of the same event can be run concurrently~\cite{jones:2017}.
\item Individual framework modules are allowed to
parallelize their work with multi-threading using Intel Threading Building Blocks (TBB) primitives~\cite{TBB}, or
with vectorization.
\begin{itemize}
  \item Only a few modules currently attempt parallelization with TBB.
  \item Vectorization level in production is still SSE3. Nevertheless, an option for building the same library for multiple vector architectures and picking the proper one at run time is supported, although not used in production yet. 
\end{itemize}
\item Data from different events can be processed concurrently~\cite{jones:2014,jones:2015}.
\begin{itemize}
    \item Events from different recorded data units (``LuminosityBlocks'') can be processed concurrently~\cite{Jones:2018dti}.
    \item Events from different Intervals-of-Validity (IOV) of calibrations can be processed concurrently~\cite{Dagenhart:2020igp}.
    \item Support for processing Events from different Runs concurrently is being worked on. This development leaves the processing of Events from different input files as the remaining synchronization point.
\end{itemize}
\end{itemize}
Note that some of these features impose certain requirements on the framework module code, and therefore the extent to which specific applications (like simulation, reconstruction, or data quality monitoring) make use of those, depends on the application. For instance, in order to process different IOVs concurrently, the module code needs to avoid caching the IOV-dependent objects it uses and instead passes them as arguments to the relevant function calls; modules that do not comply with this requirement cannot be used to process different IOVs concurrently, but may still be able to exploit other parallelization features enabled by the framework.

The CMSSW framework has a generic ability to ``offload" work to any external processing system~\cite{bocci:2020}. The offloading is asynchronous, i.e. the CPU thread offloading the work can continue to work on other tasks while the offloaded work is being run elsewhere. Using this generic mechanism offloading to NVIDIA GPUs on the same node as the CPU process is supported. Note that strictly speaking this support is not part of the framework itself, but it is an additional layer between the framework and the user code. In addition, the Services for Optimized Network Inference on Coprocessors (SONIC)~\cite{SONICgpu,SONICfpga} approach for remote ML inference has been integrated in CMSSW and its usage is being considered for CMS in HL-LHC. 
CMSSW is also phasing in support for portable code implementations with Alpaka~\cite{alpaka}, allowing for portability across CPU and NVIDIA GPUs; this initial support can be extended to other tools (such as those discussed in Sec.~\ref{sec:portability}) and to support AMD and Intel GPUs.

\section{Experimental Algorithms and their Connection to AI/ML: Challenges and Opportunities}

AI/ML is playing an increasingly important role in HEP and could be a significant resource when developing experimental algorithms. AI/ML methods, such as Graph Neural Networks~\cite{graphNN}, have been studied for track reconstruction, object identification, and pileup rejections at LHC experiments, reconstruction at NOvA,
and at cosmic frontier experiments.
Not only can AI/ML improve on current experimental algorithms, but it can also serve as a portability solution.
The AI/ML frameworks have in the past supported both CPUs and GPUs and are expected to support GPUs from multiple vendors (e.g., NVIDIA, Intel, AMD, etc). Thus, AI/ML algorithms together with common AI/ML frameworks, e.g., TensorFlow~\cite{tensorflow2015-whitepaper}, PyTorch~\cite{NEURIPS2019_9015}, could be used as a surrogate to allow experimental algorithms to run on a large range of different hardware.

Additionally, certain experimental algorithms may require fast inference (e.g., in LHC trigger applications). This is only a challenge if a limited buffer is available, making the batch size for inference small. Work is ongoing to address this~\cite{needsToolsResources} and could expand the use of AI/ML as a surrogate for more experimental algorithms.

The use of AI/ML, however, comes with challenges. Traditional non-AI/ML experimental algorithms are usually transparent and can be easily interpreted as to what physics knowledge and information is used to achieve a certain goal. AI/ML algorithms can be black boxes for which the inner workings are difficult to interpret which in turn could lead to unexpected results. Additionally, AI/ML can be sensitive to the training set that is used and may require significantly more data to optimize than traditional physics-inspired algorithms which make use of physicists' knowledge (e.g., Lorentz symmetries, known detector features/limitations). Thus AI/ML surrogate experimental algorithms need to both better incorporate domain knowledge for more efficient and robust learning but also need to be validated to ensure high-quality outputs of the algorithms.

Finally, the use of common ML frameworks and algorithms with similar structures, e.g., convolutional neural networks, could offer an opportunity to co-design hardware and software with chip vendors that most closely meets the demands of experiments. These designs would need to be usable across experiments/frontiers given the significant investment needed for a more application specific hardware design. 

\section{Algorithms in the Experimental Frontiers}
\label{sec:algsfronts}

So far, we have focused the discussion on opportunities and challenges with regard to experimental algorithms that exist across the frontiers. Next we provide an overview of notable algorithms and developments, split by physics frontier.

\subsection{Energy Frontier}
The energy frontier (EF), with the start of the High-Luminosity (HL) LHC, will be entering an era where unprecedented data volumes 
need to be processed and stored. Future experiments will not only have their data volumes increase but will also increase in complexity which will require novel algorithms to extract their full discovery potential. Finally, the changing computing architecture landscape will require significant changes in how experimental algorithms are implemented.

Experimental algorithms within the EF consist of algorithms that reconstruct physics objects, e.g., electrons, muons, the products from quarks (jets), from either detector simulations or detector readouts. Additionally, these algorithms can be classified as ``online'', which are part of the data taking process and require low latency due to the rate of collisions, and ``offline'' which are run on already stored data, are more refined, and are not bound to the same time constraints as online. These reconstruction algorithms are developed by many different experts with greatly varying expertise, i.e., in a particular detector subsystem, and have been optimized for use on x86-based CPUs. Due to the sheer number of algorithms and the limited person power with the required expertise, both in the subsystem and software engineering, portability between different platforms and vendors is absolutely necessary for future experimental algorithms.

The main challenge for future collider experimental algorithms is to maintain the current physics performance in spite of the significant increase of pileup (i.e. the number of proton-proton interactions per bunch crossing, going from 20-60 at the LHC to 140-200 at the HL-LHC). There are many reconstruction algorithms but charged particle trajectory reconstruction (tracking) requires significant computational improvements for the HL-LHC~\cite{HEPSoftwareFoundation:2020daq} and beyond. In fact, tracking algorithms are typically based on combinatorial searches whose complexity and timing scales poorly (i.e. quadratically or worse) with increased detector occupancy as those resulting from the HL-LHC pileup. The HEP community has already made progress in improving tracking and preparing to make use of more parallelism and heterogeneous computing both with AI/ML (e.g., Exa.TrkX~\cite{Ju:2020xty}) and without AI/ML (e.g., ACTS~\cite{Ai:2019kze,Gumpert_2017}, mkFit~\cite{Lantz:2020yqe}, Patatrack~\cite{andrea_bocci_2019_3598824}, and Allen~\cite{Aaij:2019zbu}). These efforts not only need to be brought to maturity to have experimental algorithms ready for the HL-LHC and other future colliders but person power also needs to be allocated to maintain and improve the algorithms.

In addition to ensuring that the performance of reconstruction algorithms remain relatively constant, EF experiments also require improved triggers which can capture interesting events in a high pileup environment. These online algorithms are typically cruder and faster versions of offline reconstructions algorithms. There are efforts to make the online algorithms be more similar to the offline reconstruction algorithms. This would improve the sensitivity of future physics analyses by allowing for finer grained information, e.g., using the full calorimeter granularity rather than collections of them, to be used at the trigger level which in turn would result in higher resolution for physics object properties used for triggering (e.g., jet transverse momentum). An additional benefit to having closely related online and offline reconstruction algorithms is that both systems could benefit from heterogeneous computing resources with minimal rewriting of code.

To maintain physics performance in future EF experiments, which will be more complex and see significantly higher data volumes, experimental algorithms need to be adapted for a heterogeneous computing environment. Due to the complexity of the experiments and due to budget and computing constraints, future algorithms need to be portable so they can be run on multiple processor types.

\subsection{Cosmic Frontier}

The field of cosmology has undergone a major transformation over the last two decades due to the advent of large sky surveys that provide detailed information about the make-up and evolution of the Universe. The data sets collected by the surveys have required a sea change in the development of experimental algorithms in the field, spanning data processing algorithms as well as data analysis algorithms. For optical surveys, the algorithms can be broadly classified as falling into two classes: algorithms for static science where the data is collected over a long span of time, usually at least a year, and time domain science, where algorithms are needed to find events quickly, classify them and provide guidance for further follow-up actions. In both cases, AI/ML algorithms and heterogeneous computing can play an important role. Ref.~\cite{2019arXiv190505116B} provides an excellent overview of the challenges and opportunities with regard to algorithm and software development. 

In addition to the image processing algorithms, substantial analysis software infrastructure must be developed and demonstrated to work robustly and at scale: key algorithms and code needed to extract knowledge from the data, frameworks to enable petabyte-scale analyses, and mechanisms to search for one-in-a-million or one-in-a-billion events in continuous streams of data.  Fortunately, diverse science objectives of surveys like Rubin Observatory's Legacy Survey of Space and Time (LSST) share many common traits and computational challenges. Algorithms and analysis frameworks developed to answer one particular question can, if designed with forethought and using professional software engineering practices and industry tools where appropriate, enable many core science objectives. Doing so requires resources and community support for software development that is coordinated across individuals with a range of levels of computing expertise, from the traditional groups of scientists to computational scientists, including dedicated long-term personnel. Moreover, development of algorithms and approaches to effectively make use of the evolution of computing architecture at supercomputing facilities (e.g., adopting algorithms that are amenable to GPU acceleration) requires some forethought and support.

The algorithmic challenges for Cosmic Microwave Background (CMB) experiments are also growing rapidly. In particular with the advent of CMB-S4 in the late 2020s, data volumes will increase more than an order of magnitude from current experiments, reaching 100-petabyte-scale data volumes by the end of the survey ($\sim 50 \mathrm{TB}/\mathrm{day}$). The primary computational challenge is reduction of these raw detector data to intensity and polarization maps, and then simulation of the instrument and map-making strategy. Handling these unprecedented data volumes will require substantial new development of software and stress data-transport and storage systems given the relatively high data volume per cycle of most CMB data processing. Minimizing this computational load will require use of computational accelerators, historically rarely used in CMB analyses because of data-throughput limitations and strongly heterogeneous manipulations not dominated by a single operation, organizational efforts to minimize duplication of data processing, and exploitation of a variety of computing systems, including supercomputing centers, distributed resources, and field computation, for example at the South Pole. Some, though not all, of the algorithms currently used in the field are embarrassingly parallel and able to run on widely distributed, shared-nothing systems, while others require tightly-coupled architectures operating on large fractions of the instrument data; a mixture of these will likely be involved in CMB-S4 data processing.

Direct dark matter detection experiments face a different set of challenges with regard to experimental algorithms than the cosmological surveys. The Snowmass White Paper~\cite{Kahn:2022kae} on {\em ``Modeling, statistics, simulations, and computing needs for direct dark matter detection''} provides an excellent overview. They state that their current experiments are approaching data volumes of $\sim$1 PB/year, see, e.g., Ref.~\cite{Mount:2017qzi}. This poses a significant challenge since the development of a scalable computing infrastructure has not been a priority for the experiments so far. They report challenges specifically with regard to data management and archiving, event processing, reconstruction and analysis, software management, validation and distribution. In the context if this Snowmass report, event processing and reconstruction are particularly relevant. Their challenges closely mirror those of particle physics experiments. They discuss the evolution of computing models in the view of HPC resources and the usage of scalable HEP frameworks, a topic central to this report. The White Paper also discusses the use of machine learning techniques and point out early success but also remaining challenges. The importance of cross-discipline collaborations is highlighted. Finally, they stress the importance of workforce development, a topic highlighted in basically every overview on computing in HEP experiments.

\subsection{Neutrino Frontier}

Experimental algorithms for neutrino experiments, and in particular for liquid argon time projection chambers (LArTPC) detectors, can be divided in two major categories: signal processing and high-level algorithms. Signal processing algorithms include several steps operating on the waveform signal from the TPC: noise filtering, deconvolution, hit (or region of interest) finding. Higher level algorithms rely on the signal processing for physics object clustering and identification, including tracking, calorimetry, and particle identification tasks. Different methods are being employed by experiments or explored by R\&D projects, using techniques which range from traditional algorithms to deep learning networks.

Large neutrino detectors such as DUNE produce a large amount of raw data for a full event readout, imposing stringent limits on memory usage by data processing applications (as well as disk space).  
In addition, they require long processing times for both simulation and reconstruction tasks. 
The largest requirements on data processing are imposed by the need to be alert to supernova neutrino burst events~\cite{DUNE:2020ypp}, and quickly transfer and process the data in order to provide a pointing signal to optical astronomers. For example, a supernova event readout of all four DUNE modules after data compression will be of order 184 TB in size and will take a minimum of 4 hrs to transfer over a 100 Gb/s network. It will then take about 130,000 CPU-hrs for signal processing at present speeds so that, requiring processing to take the same time as transfer, a peak of 30,000 cores would be needed~\cite{Schellman:2020vdz}.
Signal processing algorithms operate on the waveform data read out from the detector. The regular data format and the modular structure of large homogeneous detectors present the opportunity of large speedups from parallelization at multiple levels (module, TPC, plane, wire), possibly exploiting accelerated computing hardware.

Challenges for DUNE data processing are described in a Snowmass White Paper using the Wire-Cell toolkit as test case application~\cite{Fleming:2022nvv}. Wire-Cell~\cite{Qian:2018qbv} is a reconstruction technique for wire-based LArTPC detectors based on a tomographic approach for native 3D event reconstruction. New parallel processing patterns are being and must continue to be employed at many levels of scale: loop level, task level, data flow level, between processes and between host computers. Jobs above a certain scale must become a distributed collection of heterogeneous parts. Distribution paths include file input/output, inter-graph data flow and GPU task servicing. Though introducing multi-threaded execution solves problems from memory limitations of HTC (grid) nodes, a new one arises due to competing thread pools and corresponding stalls. The pattern of these problems are seen both in use of CPU-only resources and with mixed CPU-GPU systems. Flexible ways to configure jobs to match the limitations of any given hardware allocation are needed. From its start, the Wire-Cell Toolkit was designed to solve some of these problems. It allows for highly multi-threaded operation and works in a synergistic manner with any TBB based application such as those based on the {\em art} event processing framework. In particular, the {\em art} multi-threading support provides potentially well matched solutions when coupled with the Wire-Cell Toolkit.

Optimization for specific algorithms is also a viable solution for neutrino experiments. The hit finder algorithm has been optimized for execution on modern CPU architecture, exploiting both vectorization and multi-threading~\cite{Berkman:2021ffy}. This development has been integrated in LArSoft, the shared software repository for LArTPC experiments (see Sec.\ref{sec:comtools}), and is being used in production by multiple experiments. Although the impact on the total processing time is small, this demonstrates the potential for large speedups using parallel computing techniques for LArTPC data processing.

Continued support for common repositories such as LArSoft is vital to enable direct sharing of developments across currently operating LArTPC experiments~\cite{FASER:2022yqp}, and to guarantee long term maintenance of such developments towards utilization in DUNE. Integration of algorithms in a common repository, as well as solutions for efficient usage of different architectures as well as HPC resources, guarantees that these tools are usable and maintained over time potentially beyond the lifetime of R\&D projects, at least to the extent that the core team (e.g. SciSoft team maintaining {\em art} and LArSoft) and the experiments are able to do so.

Workflows for running data processing tasks for neutrino experiments at HPC centers have been successfully developed, and new approaches are being developed. Early applications relied on generic container solutions so that central software builds could be used on the HPC nodes. While successful in terms of excellent scaling over multiple nodes, the main limitation of these approaches is that they were not able to exploit optimization features at node level, such as vectorization and multi-threading. New developments~\cite{LOIHEPrecoHPC} are exploring the utilization of native builds for a specific HPC center using Spack~\cite{Spack} as a build system that can allow for specific compilation of modules that can be optimized for the compute node architecture. These approaches are also exploring new tools for workflow management and object-store based data storage.

\subsection{Rare Processes and Precision Frontier}

Allen~\cite{Aaij:2019zbu} is a general framework for heterogeneous processing, whose first concrete implementation is the first-level trigger of LHCb---the first high-throughput trigger fully executed on GPUs in an HEP experiment.
Allen is a natively cross-architecture framework in which developers write CUDA/C$++$ and a thin portability layer then reinterprets the CUDA-specific parts as required. 
Compilation for x86 and AMD architectures is currently supported, while support for SYCL/OneAPI is work in progress. 
Allen achieves good compatibility of physics results across different architectures, generally at the per-mille level or better, and is fast enough when compiled for x86 so as to introduce no visible overhead in LHCb’s offline data processing.
Algorithms implemented in Allen include finding the trajectories of charged particles, finding proton-proton collision points, identifying particles as hadrons or leptons, and finding the displaced decay vertices of long-lived particles. 
Allen is currently used in Run3 data taking~\cite{LHCb:2021kxm} as first trigger step, where GPU cards connected to event builder nodes perform the initial event reconstruction and selection. 
Adopting Allen for its first trigger stage led to significant architectural simplifications of the LHCb online system, and the cost savings of the reduced networking requirements was more than the cost of the GPUs that run Allen. 
In addition, choosing Allen was motivated by an expected more rapid increase in performance for GPUs, which represents an opportunity to do more physics in Run 3 and beyond. Plans for GPU-based triggers are being explored also for other experiments in the rare events frontier, such as Mu2e-II~\cite{LOIMu2eGPU}.

Experiments in the Rare Processes and Precision Frontier (RPF) rely on massive data samples to search for suppressed or forbidden processes. Working with ever growing data sets with similar or fewer resources presents significant challenges for these experiments. The Belle II experiment is expected to collect 50 ab$^{-1}$ of data, about 50 times that collected at the first generation B-factories, Belle and BaBar, through the 2030s. Experimental algorithms for calibration, processing, storage, and analysis of these data sets will require careful planning and execution. The Belle II software~\cite{Kuhr:2018lps} uses a declarative style that is promising for accurately capturing high-level concepts and factorizing them from low-level implementations, hiding the implementation details and allowing rapid prototyping as underlying computing architectures (e.g.  GPUs) and computing models (e.g. the elimination of intermediate data products) evolve. The software was designed to enable parallel processing via forking processes, thereby avoiding issues related to memory sharing and preventing the need for software developers to understand how to write thread-safe code. Effective implementation of parallel processing has not yet been achieved and may require additional effort, particularly to achieve the proposed scalability. To handle the massive data sets expected at Belle II, the collaboration leverages heterogeneous, distributed computing resources using a computing model based on those of the World-wide LHC Computing Grid (WLCG) and the LHC experiments. While centralized software and tools are well supported in the field, experiments like Belle II rely on a core team of distributed computing experts who monitor the diverse computing resources, develop experiment-specific user tools, and otherwise enable efficient processing and analysis of data sets on the grid. Training and mentorship of new experts is essential to the efficient use of grid resources that enable the physics programs of RPF experiments.

\section{Recommendations}

\begin{itemize}

   
    
 

    \item {\bf Heterogeneous computing platforms} provide opportunities for significantly increased efficiency and performance for experimental algorithm deployment, provided certain challenges can be overcome. To ensure the continued success of our experimental programs and avoid steeply rising costs for computational infrastructure, workable solutions must be found to exploit these resources efficiently. This will require increased investment in software development. Targeted optimizations of key algorithms for experiments will likely give the largest speedups so they need to be supported until deployment in the experiment's workflows.

    \item {\bf Portability solutions} are of vital importance to the broad adoption of heterogeneous computing platforms, and it is imperative to provide continued support to develop and evaluate such solutions. These solutions must fulfill certain metrics such as ease of learning and long-term sustainability. Standardized approaches for portability (even in C++ standard) will provide an important step in this direction and should be supported.

    \item {\bf Software frameworks and common tools} play a key role in enabling experimental data processing both in large and small experiments. Software frameworks provide the infrastructure and the programming model for the implementation and execution of algorithms on the experimental data. Common tools are community-authored resources providing algorithm implementations that are applicable to multiple experiments, and often represent the most cost-effective solution. Support for continuing operation and evolution of these resources to adapt to the new computing landscape is critical for enabling the usage of parallel algorithms in production environments of the experiments.
    
    \item {\bf Interdisciplinary collaborations and programs} are critical to develop the most impactful algorithmic approaches that can take full advantage of new hardware and software developments. The support of such programs needs to be strengthened considerably and will be extremely valuable for the HEP community.
    
    \item {\bf Training opportunities} 
    are essential to ensure that the experiments implement optimal algorithms that can take full advantage of new hardware and software developments. The computational landscape has become very complex in the last few years and extensive training is essential to successfully navigate it. Standardized approaches for portability (even in C++ standard) may lower the bar for training/specialization of workforce

    \item {\bf Career opportunities} need to be provided for researchers who focus on experimental algorithm development. The field has become very complex and requires deep knowledge of diverse computational approaches. The High Energy Physics community often does not recognize the scientific value in this work when faculty positions are competed; scientific research positions at university have almost disappeared. While there are opportunities at the national laboratories, they are clearly too few and the field is losing very talented researchers to industry at an alarming rate.
    
    \item {\bf Project support} has to be provided to enable long-term sustainability of software developments and to offer job security for early-career researchers. In a similar way that detector development is supported by projects, software development has to be treated in a similar way. Establishing synergies between detector design and data processing algorithm design will likely provide optimal physics results and overall cost effectiveness.

    \item {\bf Resource allocations} need to be sufficient to go beyond the R\&D phase of algorithm and software development to production ready software. Often, funding is provided to demonstrate that a new idea or algorithm in principle works, however, follow-up funding to take the very important step to go from proof-of-concept to fully production ready and maintenance is often not available. The funding agencies have to consider long-term commitment for mission-critical software products. 
\end{itemize}





\bibliographystyle{JHEP}
\bibliography{Computation/CompF01/myreferences}

\end{document}